\documentclass[twocolumn]{jpsj2} 
\title{Electronic Structures of CaAlSi with Different Stacking AlSi Layers by First-Principles Calculations}

\author{Sogo \textsc{Kuroiwa}$^1$, Akiyoshi \textsc{Nakashima}$^1$, Shin \textsc{Miyahara}$^{1,2}$, Nobuo \textsc{Furukawa}$^{1,2}$ and Jun \textsc{Akimitsu}$^1$}

\inst{$^1$Department of Physics and Mathematics, Aoyama-Gakuin University, Fuchinobe 5-10-1, Sagamihara, Kanagawa 229-8558, Japan \\
$^2$Multiferroics Project (MF), ERATO, Japan Science and Technology Agency (JST), c$/$o Department of Applied Physics, The University of Tokyo, 7-3-1 Hongo, Bunkyo-ku, Tokyo 113-8656, Japan \\}

\abst{
The full-potential linear augmented plane-wave calculations have been applied to investigate the systematic change of electronic structures in CaAlSi due to different stacking sequences of AlSi layers.
The present $ab$-$initio$ calculations have revealed that the multistacking, buckling and 60$^{\rm \circ}$ rotation of AlSi layer affect the electronic band structure in this system.
In particular, such a structural perturbation gives rise to the disconnected and cylindrical Fermi surface along the M-L lines of the hexagonal Brillouin zone.
This means that multistacked CaAlSi with the buckling AlSi layers increases degree of two-dimensional electronic characters, and it gives us qualitative understanding for the quite different upper critical field anisotropy between specimens with and without superstructure as reported previously.
}

\kword{CaAlSi, superstructure, electronic properties, Fermi surface, first-principles calculations}

\begin{document}
\maketitle


A ternary silicide superconductor CaAlSi with the superconducting transition temperature ($T_{\rm c}$) $\sim$ 8 K is attracting much interest, as they exhibit an isostructural to MgB$_2$ \cite{nagamatsu} and thereby they may serve as one of the reference system to understand the high-$T_{\rm c}$ of MgB$_2$ \cite{imai}.
As a peculiar feature, CaAlSi has two types of multi-stacked structure, where they possess a clear five-layered (5$H$-CaAlSi, $T_{\rm c}$ $\sim$ 5.7 K) or six-layered (6$H$-CaAlSi, $T_{\rm c}$ $\sim$ 7.8 K) superlattice along the $c$-axis caused by the buckling and$/$or rotation of AlSi layer \cite{sagayama}.
Meanwhile, we have reported that the specimen without superstructure (1$H$-CaAlSi, $T_{\rm c}$ $\sim$ 6.5 K) exhibits the AlB$_2$-like crystal structure with Al and Si atoms being distributed regularly in the B$_2$ plane without superlattice structure \cite{kuroiwa}.
Interestingly, we have found that $T_{\rm c}$ and many other transport properties in CaAlSi depend on the superstructured periodicity.
For instance, the field-induced magnetic response in 1$H$-CaAlSi exhibits almost isotropic characteristics for each crystal axis, while that in 5$H$- and 6$H$-CaAlSi (superstructured CaAlSi) shows anisotropic ones.
Moreover, heat capacity \cite{kuroiwa}, tunneling spectroscopy \cite{kuroiwa:tun} and angle-resolved photoemission spectroscopy (ARPES) \cite{sato} measurements suggest the weak-coupling BCS superconductor for 1$H$-CaAlSi, while superstructured CaAlSi with the buckling AlSi layer is in a strong-coupling limit.

In order to clarify the detailed electronic structures in these new and interesting systems, we have tried to map out the overall electronic  band-structures and Fermi surface properties in 1$H$-, 5$H$- and 6$H$-CaAlSi using $ab$-$initio$ analysis.
Moreover, to estimate buckling effect in this multistacking system, we performed calculations on a virtual sixfold stacking specimen without buckling layers.


\begin{figure}
\includegraphics[width=1\linewidth]{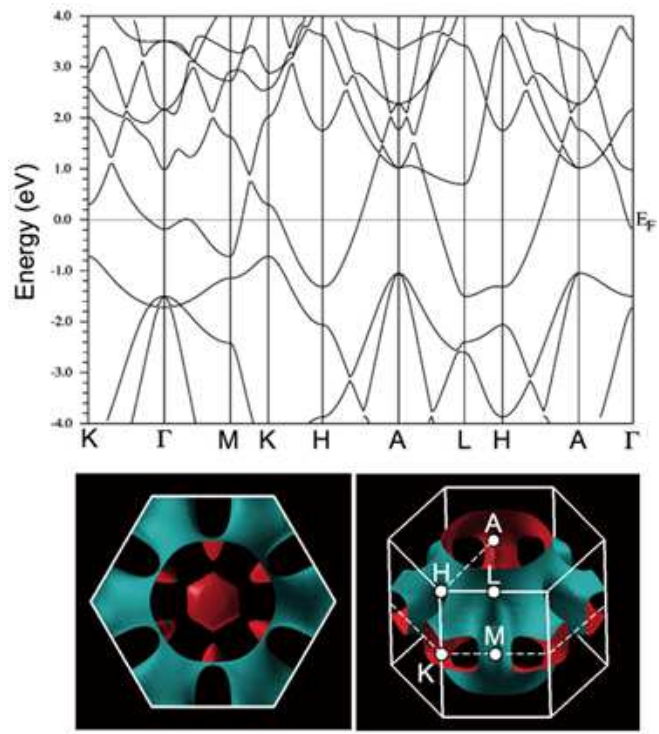}
\caption{\label{Fig1} Calculated band structure, Fermi surface and the Brillouin zone of 1$H$-CaAlSi ($P\bar{6}m2$ symmetry). }
\end{figure}

\begin{figure*}[!t]
\includegraphics[width=1\linewidth]{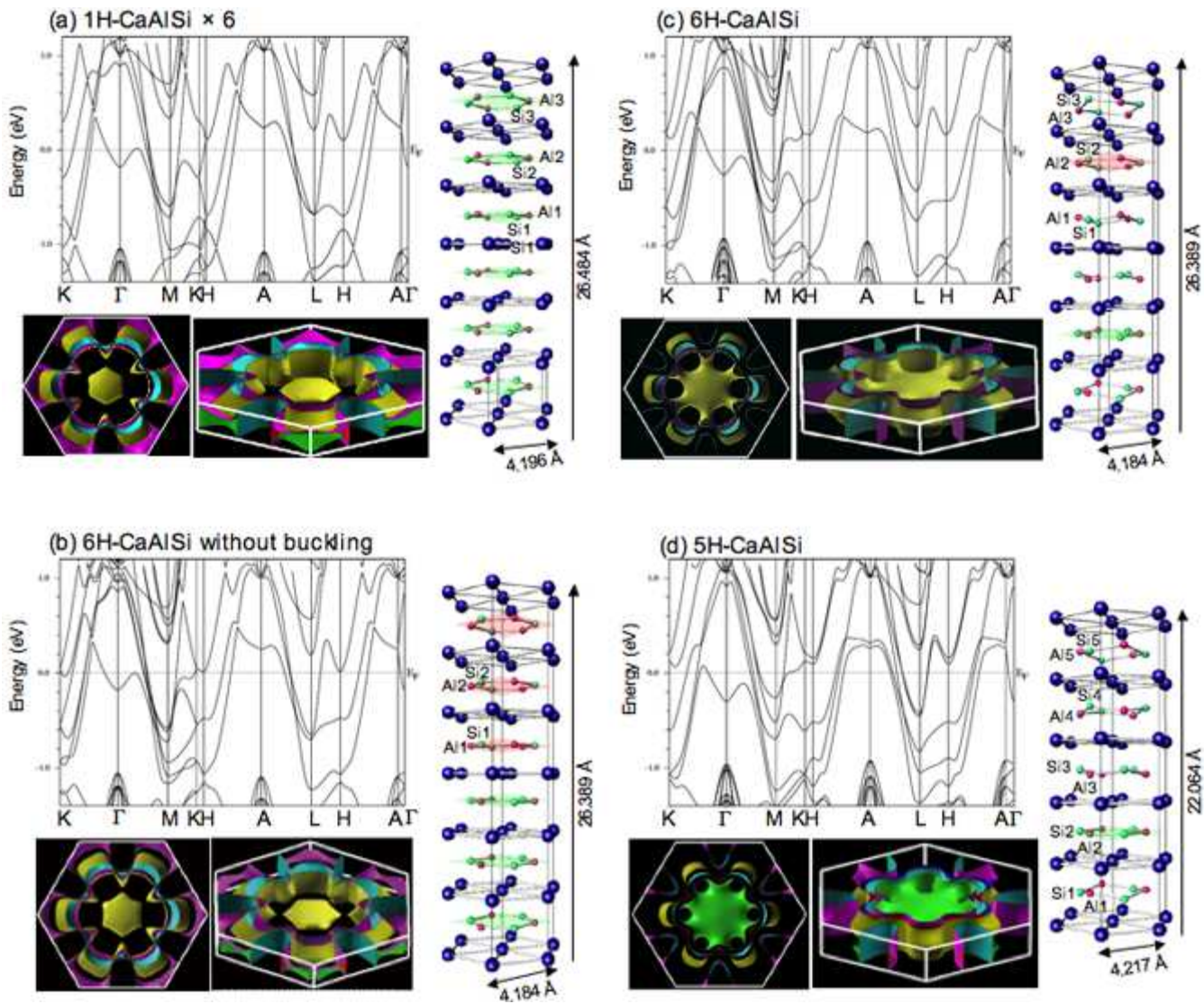}
\caption{\label{Fig2}(Color) Multistack crystal structures, calculated energy band structures and Fermi surfaces of (a) 1$H$-CaAlSi $\times$ 6 ($P\bar{6}m2$), (b) 6$H$-CaAlSi without buckling ($P6_{3}/mmc$), (c) 6$H$-CaAlSi ($P6_{3}mc$) and (d) 5$H$-CaAlSi ($P\bar{3}m1$). The green and red sheets placed into the crystal structures represent the flat AlSi layer, and their different colors mean the rotation of 60$^{\rm \circ}$ around the $c$-axis.}
\end{figure*}

Calculations of the electronic structures were performed by the full-potential linear augmented plane-wave (LAPW) method as implemented in the WIEN2K program package \cite{Blaha}.
The valence electrons were applied to a scalar-relativistic approximation, and the Perdew-Burke-Ernzerhof generalized-gradient approximation (GGA) potential was used for the exchange correlation potential \cite{Perdew}.
For the structure parameter, experimentally determined crystal symmetry and lattice constants were applied and internal coordinates were taken from the structure data reported previously \cite{sagayama,kuroiwa}.
The LAPW sphere radii of 1$H/5H/6H$-CaAlSi were set to 2.50$/2.50/2.50$ a.u. for Ca, 2.29$/2.27/2.27$ a.u. for Al and $2.29/2.27/2.27$ a.u. for Si, respectively. 
The cut-off parameter $RK_{\rm max}$ was chosen as 7.0. 

Figure {\ref{Fig1}} shows the calculated band structure, Fermi surface and the Brillouin zone of 1$H$-CaAlSi with the space group of $P\bar{6}m2$ symmetry.
We first note that our results on the overall band structures and Fermi surface properties for 1$H$-CaAlSi are in fairly well agreement with those reported previously \cite{mazin,matteo,shein,huang}. 
Unlike MgB$_2$ \cite{Kortus}, the Fermi level is mainly composed by the band of Ca $d$-orbitals: their ratio at the Fermi level is about 60$\%$ in comparison with about 10$\%$ for Al $p$-orbitals and 10$\%$ for Si $p$-orbitals.
Thus, the electronic ground state in 1$H$-CaAlSi has almost three-dimensional (3D) characteristics. 
Meanwhile, in this paper we mainly present that the electronic structure of superstructured CaAlSi increases degree of two-dimensional (2D) features induced by the structural change such as different stacking sequences of buckling and flat AlSi layers.

Figure {\ref{Fig2}} displays the multistack crystal structures, calculated energy band structures and Fermi surface of (a) 1$H$-CaAlSi $\times$ 6 ($P\bar{6}m2$), (b) 6$H$-CaAlSi without buckling ($P6_{3}/mmc$), (c) 6$H$-CaAlSi ($P6_{3}mc$) and (d) 5$H$-CaAlSi ($P\bar{3}m1$).
\par
Here, ``1$H$-CaAlSi $\times$ 6" is defined by a sixfold stacking structure of 1$H$-CaAlSi with flat AlSi layer along the $c$-axis, and then lattice parameters $a$ and $c$ were set to be 4.196 {\AA} and 26.484 (= 4.414 $\times$ 6) {\AA} as listed in our previous report \cite{kuroiwa}.
As shown in right side of Figs. 2 (c) and (d), the crystal structures of 6$H$- and 5$H$-CaAlSi possess two and one flat AlSi layers in addition to the several buckling layer, respectively.
The differences between two types of multistacked structures are attributed to the 60$^{\rm \circ}$ rotation of the Al$/$Si arrangement in the AlSi layers around the $c$-axis and the buckling and flat AlSi layers in different stacking sequences.
Structural parameters reported in Ref. [3] were applied to the present calculation on 6$H$-CaAlSi ($a$ = 4.184 {\AA} and $c$ = 26.389 {\AA}) and 5$H$-CaAlSi ($a$ = 4.217 {\AA} and $c$ = 22.064 {\AA}).
Moreover, hypothetical ``6$H$-CaAlSi without buckling" is defined by the sixfold superstructure without the buckling in the AlSi layer (i.e., assuming only flat AlSi layers) to compare with 6$H$-CaAlSi.
For 6$H$-CaAlSi without buckling, the unit cell size is set to be identical to that of real 6$H$ structure (see Fig. {\ref{Fig2}} (b)), while each flat AlSi layers are placed at the midpoint between two Ca layers.

\begin{figure}
\includegraphics[width=1\linewidth]{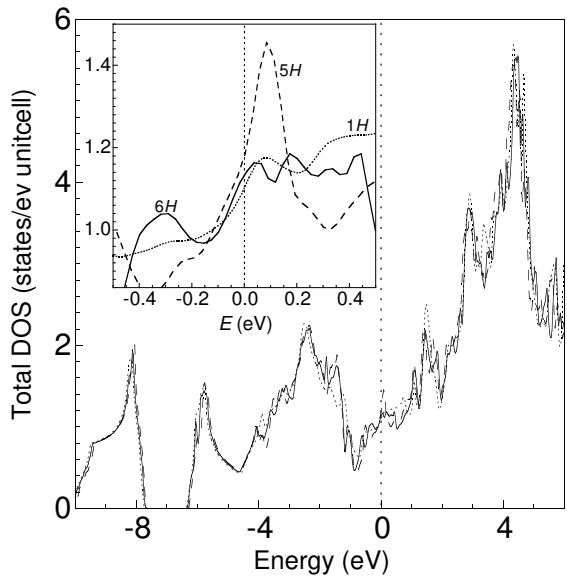}
\caption{\label{Fig3} Electronic total DOS for 1$H$-CaAlSi (dotted line), 5$H$-CaAlSi (dashed line) and 6$H$-CaAlSi (solid line). Inset shows the DOS around the Fermi level.}
\end{figure}

\begin{figure}
\includegraphics[width=1\linewidth]{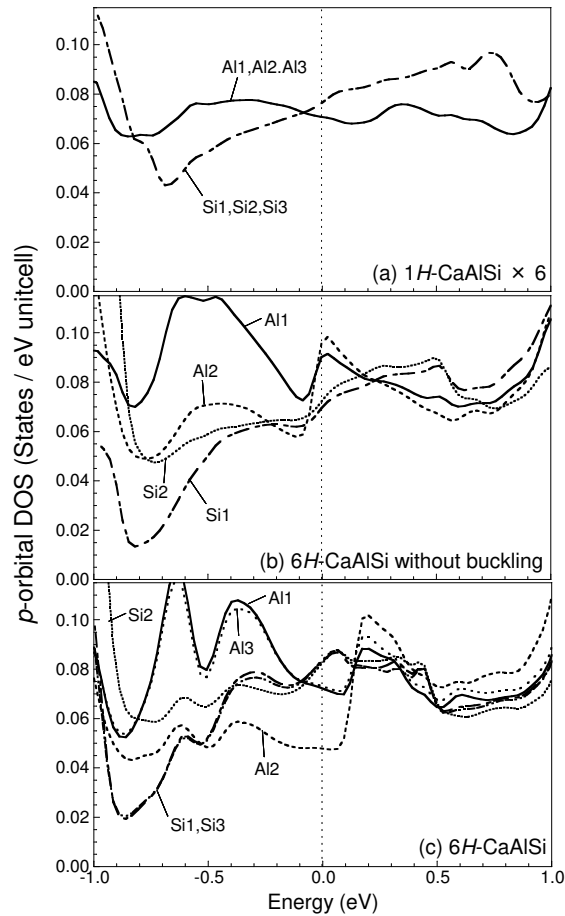}
\caption{\label{Fig4} Site-projected $p$-orbital DOS of Al and Si atoms in (a) 1$H$-CaAlSi $\times$ 6, (b) 6$H$-CaAlSi without buckling and (c) 6$H$-CaAlSi.}
\end{figure}

\par

In the following, we show that the systematic structural change in the sequence 1$H$-CaAlSi $\times$ 6 $\rightarrow$ 6$H$-CaAlSi without buckling $\rightarrow$ 6$H$-CaAlSi leads to a clear change of band structures and Fermi surface properties around K(H) points.
Comparing with the Fermi surface of 1$H$-CaAlSi $\times$ 6 and 6$H$-CaAlSi without buckling reveals the slight opening of gap at the zone boundary along the $k_z$ = $\pm{\pi}/c$ directions, due to the breaking of the translation symmetry originated from the rotation of AlSi layer.
However, no significant change of overall band structure and Fermi surface properties between them could be seen at the level of the present data.
On the other hand, in real 6$H$ structure the band lying on the Fermi energy around K-H line seems to shift toward higher energy region probably due to the additional contribution such as hybridization of Al$/$Si $p_{x+y}$- and $p_z$-orbitals because of the buckling of AlSi layer.
Thus, 6$H$-CaAlSi dose not possess any bands at the Fermi energy along the K-H lines.
In fact, the Fermi surface around K(H) points disappear in 6$H$-CaAlSi, and the disconnected and cylindrical Fermi surface sheets remain around the Brillouin zone boundary (M point) as shown in Fig. {\ref{Fig2}} (c). 
A similar relationship between the appearance of cylindrical Fermi surface and the presence of superstructure is also observed in 5$H$-CaAlSi (see Fig. {\ref{Fig2}} (d)).
Therefore, such a disconnected and cylindrical Fermi surface, which reflects to the 2D-like characteristics,  is derived from not only the contribution of Brillouin zone folding due to the multistacking structure along the $c$-axis but also the buckling effect in the AlSi layer.
\par

Figure {\ref{Fig3}} shows the calculated total DOS for 1$H$-, 5$H$- and 6$H$-CaAlSi.
Inset indicates the data near the Fermi energy, and these dotted lines at $E$ = 0 correspond to the Fermi energy.
As mentioned earlier, despite the drastic change of Fermi surface around K(H) points in superstructured CaAlSi with respect to 1$H$-CaAlSi, there is no substantial difference between the overall total DOS for these three phases.
As shown in inset of Fig. {\ref{Fig3}}, we find that total DOS at the Fermi energy slightly increases along the series 1$H$ (1.10 states$/$eV unitcell) $\rightarrow$ 6$H$ (1.13 states$/$eV unitcell) $\rightarrow$ 5$H$ (1.18 states$/$eV unitcell). 
We next discuss the origin of disappearance of the Fermi surface around K(H) points from viewpoint of site-projected orbital DOS.

\par
Figure {\ref{Fig4}} shows the site-projected $p$-orbital DOS of Al and Si atoms in (a) 1$H$-CaAlSi $\times$ 6, (b) 6$H$-CaAlSi without buckling and (c) 6$H$-CaAlSi.
Here, Al1$/$Si1, Al2$/$Si2 and Al3$/$Si3 correspond to each site of Al and Si atoms as listed in Fig. {\ref{Fig2}}.
At the level of the data here of 1$H$-CaAlSi $\times$ 6 and 6$H$-CaAlSi without buckling, there is no change of the $p$-orbital DOS at the Fermi level between each sites of Al and Si atoms, as expected for their crystal symmetry.
For 6$H$-CaAlSi, it is interesting to note that the $p$-orbital DOS of Al2 (corresponding to Al in the flat layer) at the Fermi energy is extremely smaller than that in the buckling layer (Al1 and Al3).
Therefore, such a decrease of the $p$-orbital DOS of Al2 might be connected with the disappearance of the 3D-like Fermi surface sheets around the K(H) points in the superstructured CaAlSi, and 2D-like Fermi surface appear along the M-L lines.
Unfortunately, there is no intuitive reason why the buckling effect gives rise to such a suppression of the $p$-orbitals DOS of Al2 site.

Finally, we briefly discuss the origin of a large difference in superconducting response (i.e., upper critical field anisotropy) \cite{kuroiwa} between 1$H$-CaAlSi and superstructured CaAlSi based on the present calculation results.
We stress again that 5$H$- and 6$H$-CaAlSi has a large upper critical field anisotropy, while 1$H$-CaAlSi shows an isotropic magnetic response as expected for the 3D electronic state.
As the present calculations pointed out, 3D Fermi surface properties in 1$H$-CaAlSi leads to such an isotropic superconducting response.
On the contrary, the presence of disconnected and cylindrical Fermi surface manifested in the present calculations on the superstructured CaAlSi might be given rise to such a large upper critical field anisotropy. 
In fact, this scenario is further supported by our recent ARPES measurements on the superstructured CaAlSi \cite{sato}.
It has revealed the main opening of superconducting gap on the Fermi surface with 2D-like characteristics around M(L) points.
Consequently, the origin of the different upper critical field anisotropy between 1$H$-CaAlSi and superstructured CaAlSi is relevant to the drastic change of Fermi surface properties around M(L) points from 3D character to 2D-like one.

In summary, we have provided the first information for the detailed electronic properties of CaAlSi with and without superstructure through mapping out the overall  band structures and Fermi surface.
Calculations have suggested that the structural change from 1$H$ to 6$H$ (i.e., multistacking, buckling and 60$^{\rm \circ}$ rotation of AlSi layer) leads to a clear change of band structure in this system.
In particular, the buckling effect is suppressed the $p$-orbital DOS of Al atoms in the flat layer, which might be connected with the disappearance of the Fermi surface around K-H lines.
This leads to the presence of disconnected and cylindrical Fermi surface sheet with the M-L dispersion, and it gives us the direct explanation for the experimentally determined different magnetic response (upper critical field anisotropy) between 1$H$-CaAlSi and superstructured CaAlSi.

\vspace{3mm}
\begin{acknowledgments}
We are grateful to Professor T. Oguchi for fruitful discussion and supporting present works.
This work was partly supported by the 21 st COE program, ``High-Tech Research Center" Project for Private Universities: matching fund subsidy from the Ministry of Education, Culture, Sports, Science and Technology (MEXT); 2002-2004 and the Grant-in-Aid for Japan Society for the Promotion of Science (JSPS) Fellows.
\end{acknowledgments}


\begin{thebibliography}{99}

\bibitem{nagamatsu} J. Nagamatsu, N. Nakagawa, T. Muranaka, Y. Zenitani, and J. Akimitsu: Nature (London) {\bf 410} (2001) 63.

\bibitem{imai} M. Imai, E. I-Hadi, S. Sadki, H. Abe, K. Nishida, T. Kimura, T. Sato, K. Hirata, and H. Kitazawa: Phys. Rev. B {\bf 68} (2003) 064512.

\bibitem{sagayama} H. Sagayama, Y. Wakabayashi, H. Sawa, T. Kamiyama, A. Hoshikawa, S. Harjo, K. Uozato, A. K. Ghosh, M. Tokunaga, and T. Tamegai: J. Phys. Soc. Jpn. {\bf 75} (2006) 043713.

\bibitem{kuroiwa} S. Kuroiwa, H. Sagayama, T. Kakiuchi, H. Sawa, Y. Noda, and J. Akimitsu:
Phys. Rev. B  {\bf 74} (2006) 014517.

\bibitem{kuroiwa:tun} S. Kuroiwa, T. Takasaki, T. Ekino, and J. Akimitsu:
Phys. Rev. B {\bf 76} (2007) 104508.

\bibitem{sato} T. Sato, $et$ $al$.:
unpublished (in preparation).

\bibitem{Blaha} P. Blaha, K. Schwarz, G. K. H. Madsen, K. Kvasnicka, and J. Luitz: WIEN2K, An Augmented Plane Wave + Local Orbitals Program for Calculating Crystal Properties.

\bibitem{Perdew} J. P. Perdew, K. Burke, and M. Ernzerhof: Phys. Rev. Lett. {\bf 77} (1996) 3865.

\bibitem{mazin} I. I. Mazin and D. A. Papaconstantopoulos: Phys. Rev. B  {\bf 69} (2004) 180512(R).

\bibitem{matteo} M. Giantomassi, L. Boeri, and G. B. Bachelet: Phys. Rev. B  {\bf 72} (2005) 224512.

\bibitem{shein} I. R. Shein, N. I. Medvedeva, and A. L. Ivanovskii:
 J. Phys.: Condens. Matter  {\bf 15} (2003) L541.

\bibitem{huang} G. Q. Huang, L. F. Chen, M. Liu, and D. Y. Xing:
 Phys. Rev. B  {\bf 69} (2004) 064509.

\bibitem{Kortus} J. Kortus, I. I. Mazin, K. D. Belashchenko, V. P. Antropov, and L. L. Boyer:
Phys. Rev. Lett. {\bf 86} (2001) 4656.
 
\end{thebibliography}
\end{document}